\documentclass{article}  
\usepackage{bigsky2009}
\usepackage{graphicx}
\usepackage{amssymb}
\frompage{000} \topage{000}                                              
\def\rightmark{Energy Loss and Medium Response}
\def\leftmark{M. McCumber et al.}

\title{Energy Loss and Medium Response via\\
Two Particle Correlations in Heavy Ion Collisions} 
\authors{
{Michael P. McCumber for the PHENIX Collaboration %
}\\[2.812mm]
{\normalsize
Department of Physics and Astronomy \\
SUNY Stony Brook \\ 
Stony Brook, NY, USA\\[0.2ex] 
}}
 
\abstract{
In these proceedings, pair correlations between high $p_{T}$ 
final-state particles in central and mid-central Au+Au are 
presented as a function of trigger-angle with respect to the 
reaction plane. Jet tomography results like these add another 
dimension of experimental discrimination to energy loss 
calculations and initial state geometric descriptions.  The 
observed increasing away-side suppression with respect to the 
nuclear overlap demonstrate that back-to-back production at 
high $p_{T}$ in mid-central collisions is not dominated by 
unmodified production produced tangentially to the nuclear 
overlap. Furthermore, these results present a challenge to 
theories that predict only a weak variation due to angle with 
respect to the reaction plane.
}

 
\begin{document}
 
\maketitle
\setcounter{page}{1}

\section{Introduction}\label{intro}

Away-side high $p_{T}$ pair correlations in heavy ion collisions 
made above the medium response at intermediate  $p_{T}$ can 
be used to study the nature of parton energy loss and system 
geometry. Varying the jets with reaction plane angle changes 
the path length through a medium in a fixed system, something 
that cannot be done via centrality selection. The dependence 
of the away-side yields on reaction plane angle will differ 
if the dominate source of surviving pairs results by partons 
crossing the nuclear-overlap versus models where complete 
energy loss in a large core region limits the away-side 
production to back-to-back partons tangential to the nuclear 
overlap.

This analysis uses pairs made between triggers of neutral pions 
between 4-7 GeV/c and charged hadrons between 3-4 and 4-5 GeV/c 
to construct per-trigger yield (PTY) for near- and away-side
jets. 

These pairs between final-state particles above intermediate 
$p_{T}$ have previously been shown to be dominated by jet
fragmentation~\cite{ppg083}. A study of away-side ($\Delta\phi 
\approx 180^{\circ}$) suppression by azimuthal angle with respect 
to the reaction plane ($\phi_{s}$) can be used to probe of 
energy loss characteristics and overlap geometry.  Pairs made
below $\sim4 GeV/c$ show new structures and may be related to 
medium response to the passage of fast partons or other new 
physics.  Intermediate $p_{T}$ effects have been discussed in 
detail previously \cite{ppg083}\cite{QMproc}. 

Despite being above known intermediate $p_{T}$ medium-response 
effects ($p_{T} \lesssim 4~GeV/c$), these data remain within 
transverse momenta that may contain recombination effects ($p_{T} 
\lesssim 6~GeV/c$) \cite{recomb}.

\begin{figure}[tb]
   \includegraphics[width=1.0\textwidth]{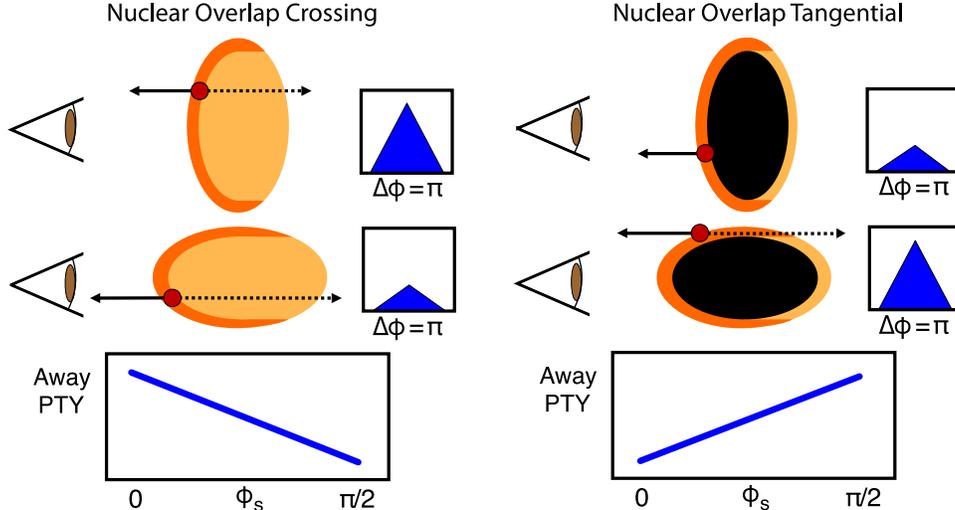}
\vspace*{-0.95cm}
\caption[]{
(Color online) Two categories of away-side jet quenching. In 
the left column, partons cross the nuclear overlap by partial 
energy loss.  In the right column, only partons tangential to 
the nuclear overlap contribute pairs in the away-side by total 
energy loss.  Both cases have a similar set of triggers.
}
\label{cartoon}
\end{figure}

Two kinds of production may contribute to the surviving 
back-to-back jet pairs. The first kind, nuclear overlap crossing 
production (see left column of Fig \ref{cartoon}), may occur 
if away-side partons lose only some fraction of their energy 
during transit or if only a fraction of away-side partons are 
completely unsuppressed.  The reaction plane dependence does not
directly differentiate how the away-side partons transits the 
nuclear overlap.  Production of the second kind is tangential 
to the nuclear overlap (see right column of Fig \ref{cartoon}) 
and occurs at some level in all cases. This production may 
dominate if energy loss of all partons entering more deeply 
into the nuclear overlap is complete and only unmodified 
tangential pairs survive.  

However, it is worth noting that models exhibiting only a small 
core region of complete energy loss may be insufficient to 
limit the surviving partons to be tangential to the nuclear 
overlap.  Surviving partons in this description may be tangential 
to the surface of the small core, but not the nuclear overlap.

Crossing production is increasingly suppressed as the path-length 
through the nuclear overlap is increased. Tangential production, 
which produces pairs as function of the integrated corona density, 
will have the opposite dependence with $\phi_{s}$.  Furthermore, 
since more triggers are produced in-plane than out-of-plane, 
PTYs in this case show suppression that decreases as a function 
of reaction plane angle simply because the fraction of triggers 
without an away-side partner has decreased.

Two sets of theoretical predictions for away-side suppression 
were available for mid-central Au+Au collisions. Thorsten Renk 
has made away-side suppression calculations for nuclear overlap 
crossing production \cite{renk}. The closest of which is for 
12-20 GeV/c triggers paired with 4-6 GeV/c partners and impact 
parameter 7.5 fm. These values show a weak dependence on reaction 
plane angle. Vlad Pantuev has also made a prediction for away-side
suppression \cite{vlad}. In this model \cite{vlad2}, a black-core formation 
time drives the transition between dominance by crossing production 
in mid-central collisions and tangential production in the
most central collisions.  In the systems with shorter overall
path-length, the crossing time becomes comparable with the 
formation time and in-plane crossing production dominates. In 
more central collisions ($<20\%$), the crossing time becomes 
long for all directions through the nuclear overlap, tangential 
production dominates, and the trend reverses. 

\section{Method}\label{method}  

\begin{figure}[tb]
\begin{minipage}{0.5\textwidth}
  \includegraphics[width=1.0\linewidth]{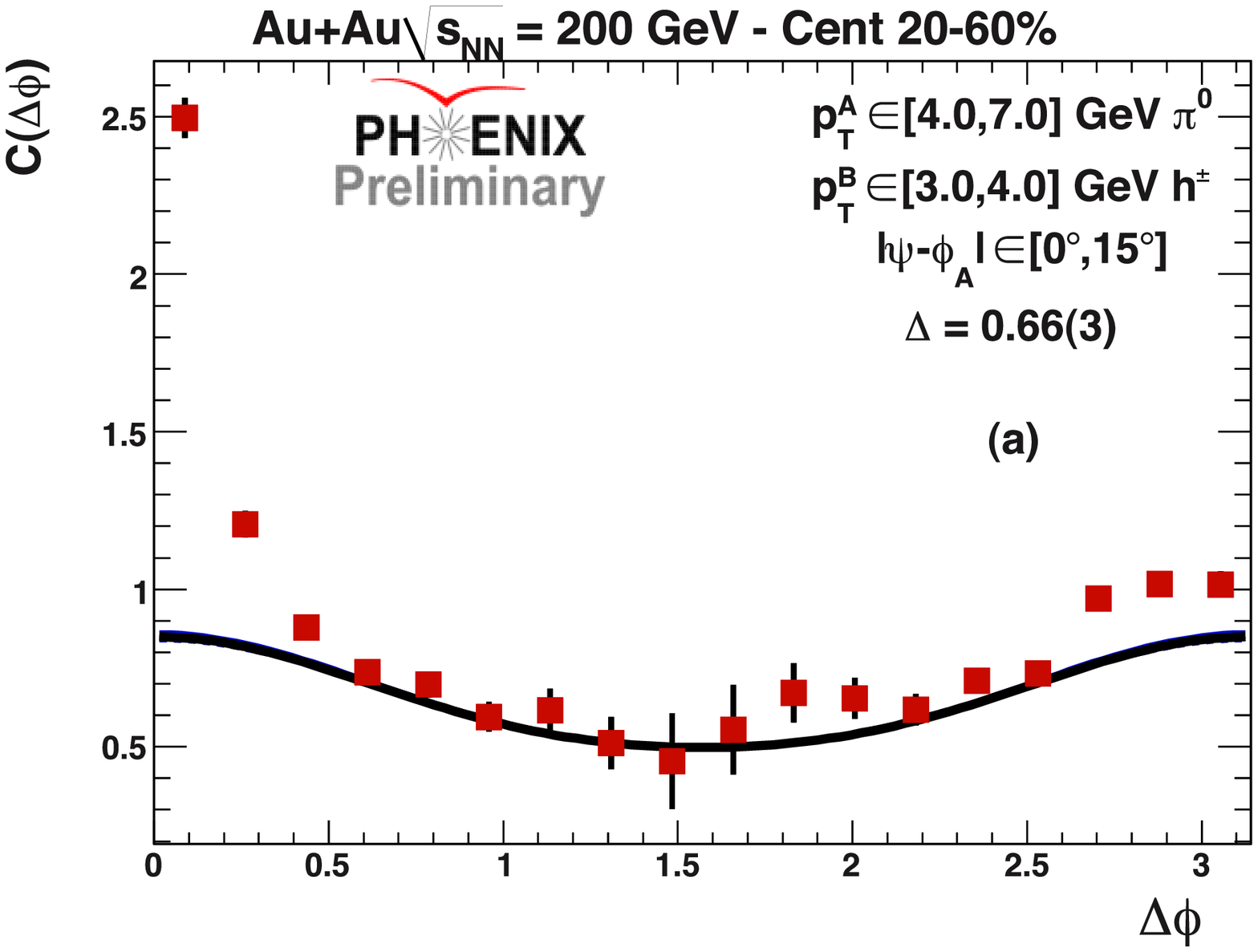}
\end{minipage}
\begin{minipage}{0.5\textwidth}
  \includegraphics[width=1.0\linewidth]{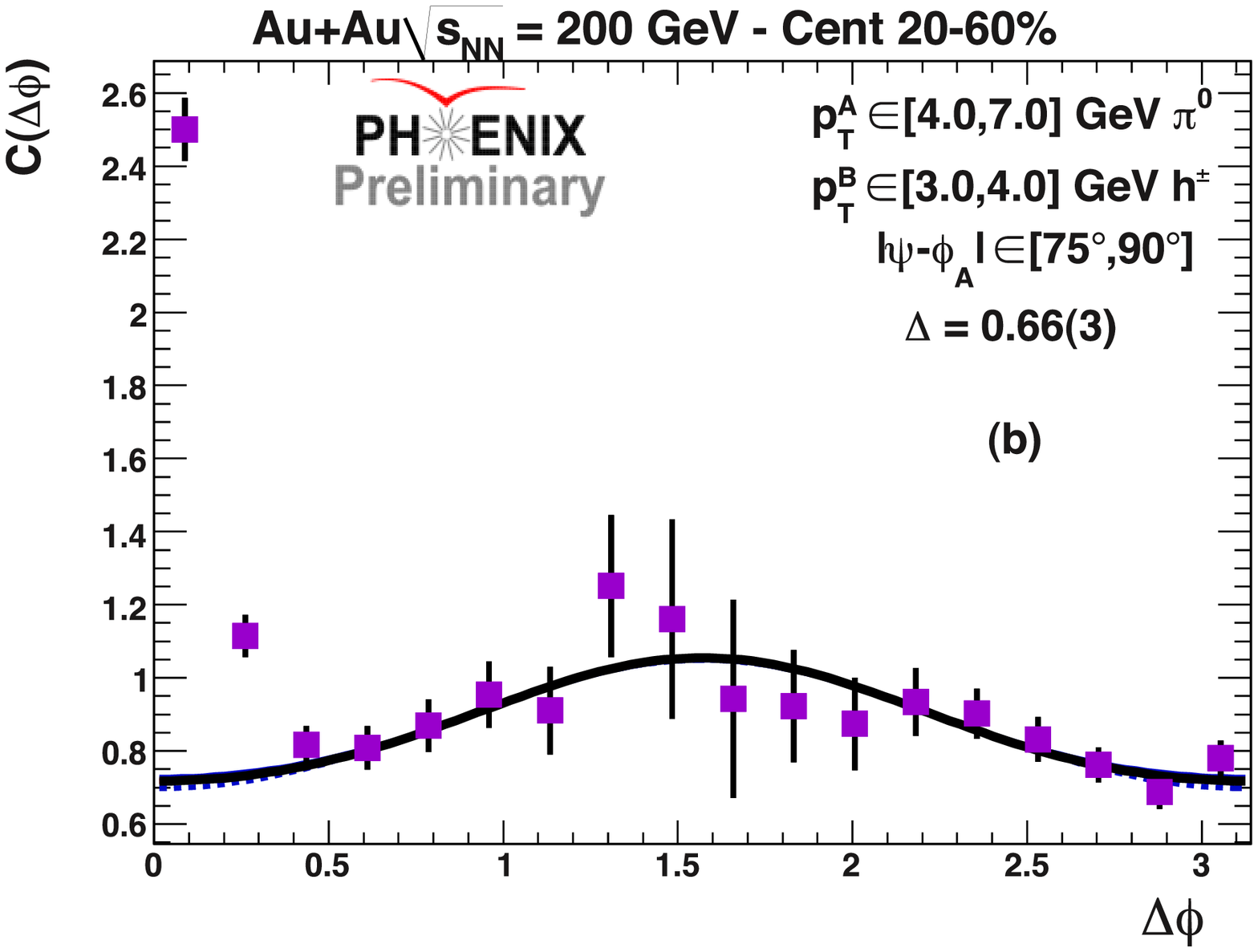}
\end{minipage}
\caption[]{
(Color online) Correlation functions depicting the two-source 
ZYAM decomposition with 3-4 GeV/c partners at mid-centrality 
for the most in-plane (a) and most out-of-plane (b) selections.
}
\label{cfs}
\end{figure}

\begin{figure}[tb]
\begin{minipage}{0.5\textwidth}
  \includegraphics[width=1.0\linewidth]{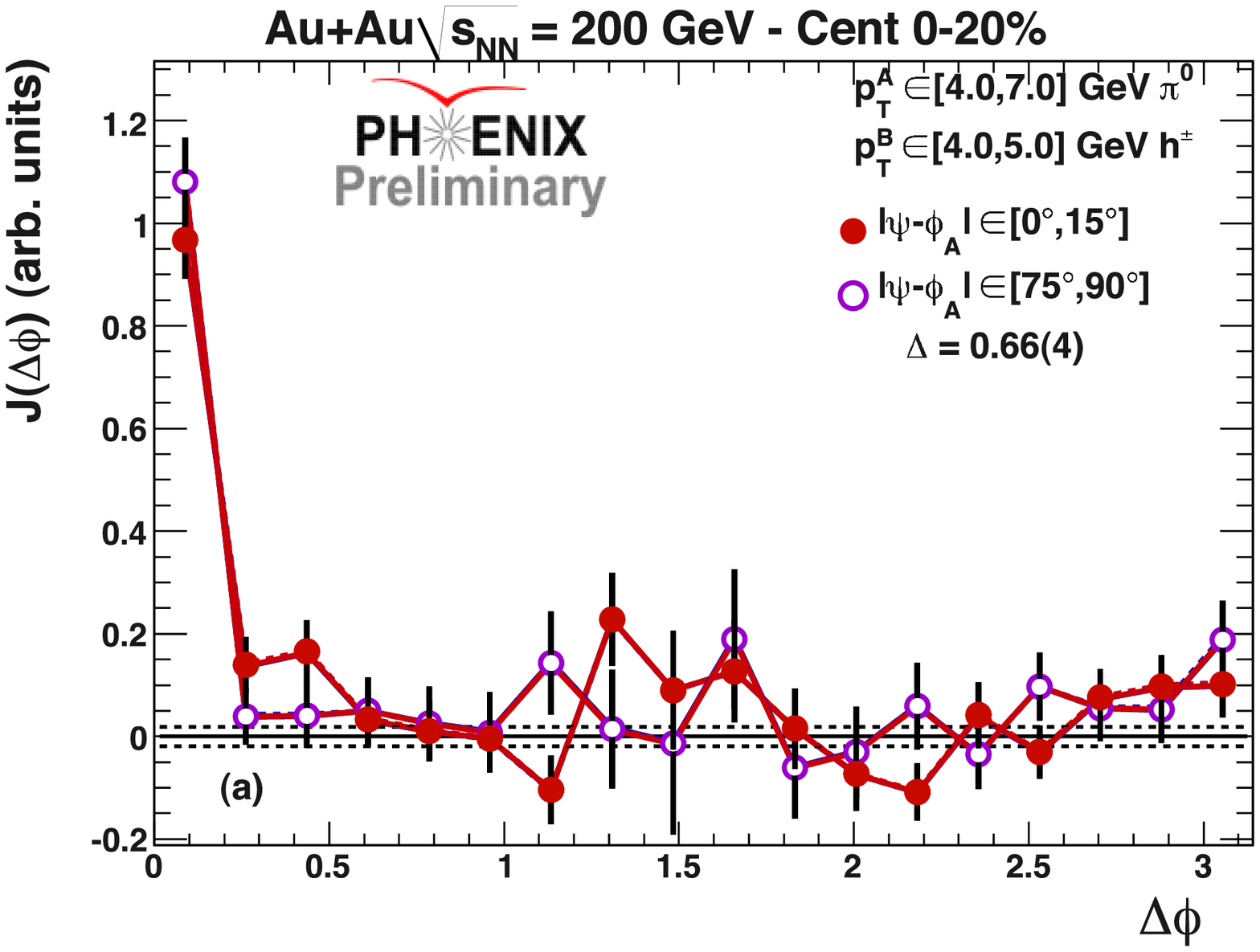}
  \includegraphics[width=1.0\linewidth]{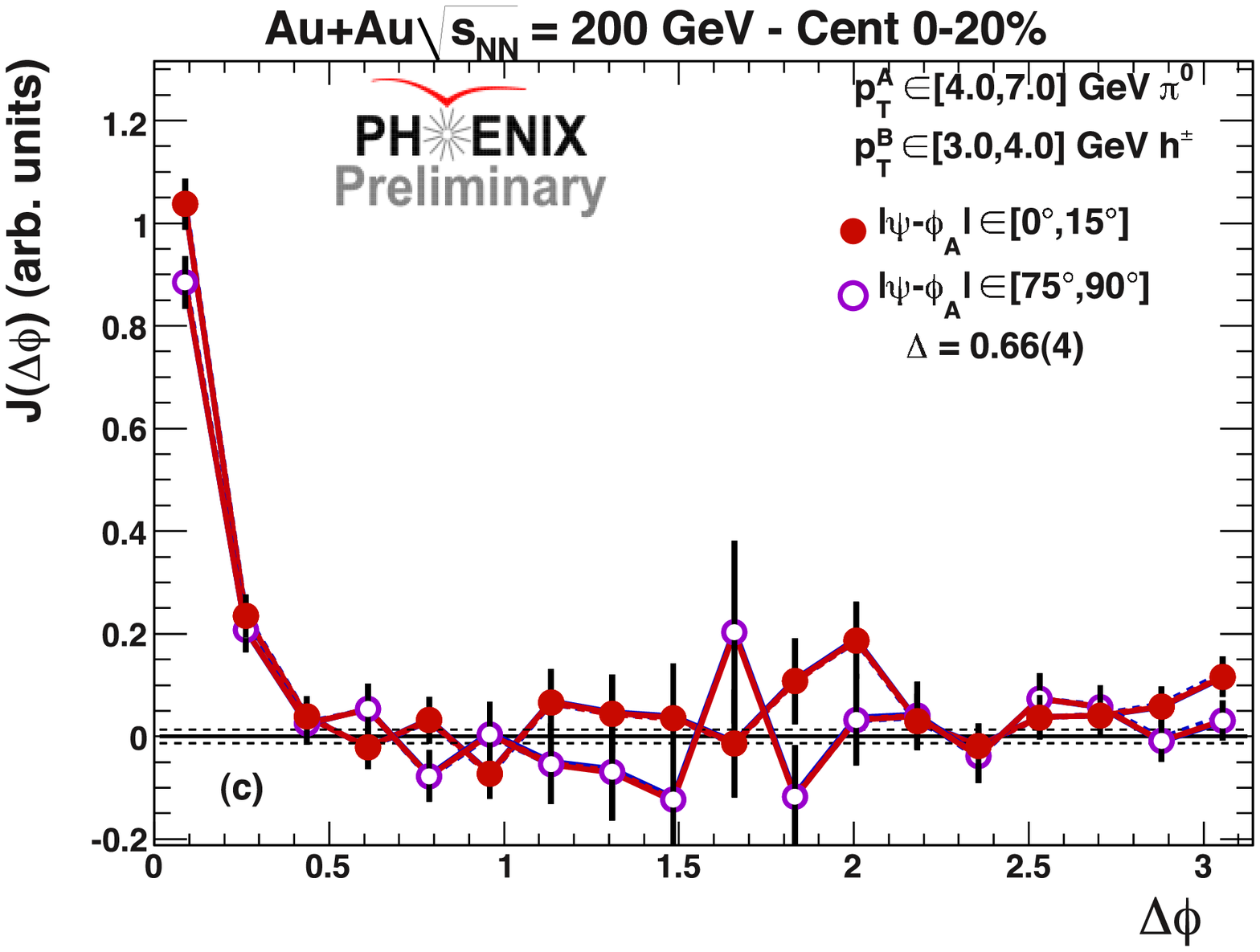}
\end{minipage}
\begin{minipage}{0.5\textwidth}
  \includegraphics[width=1.0\linewidth]{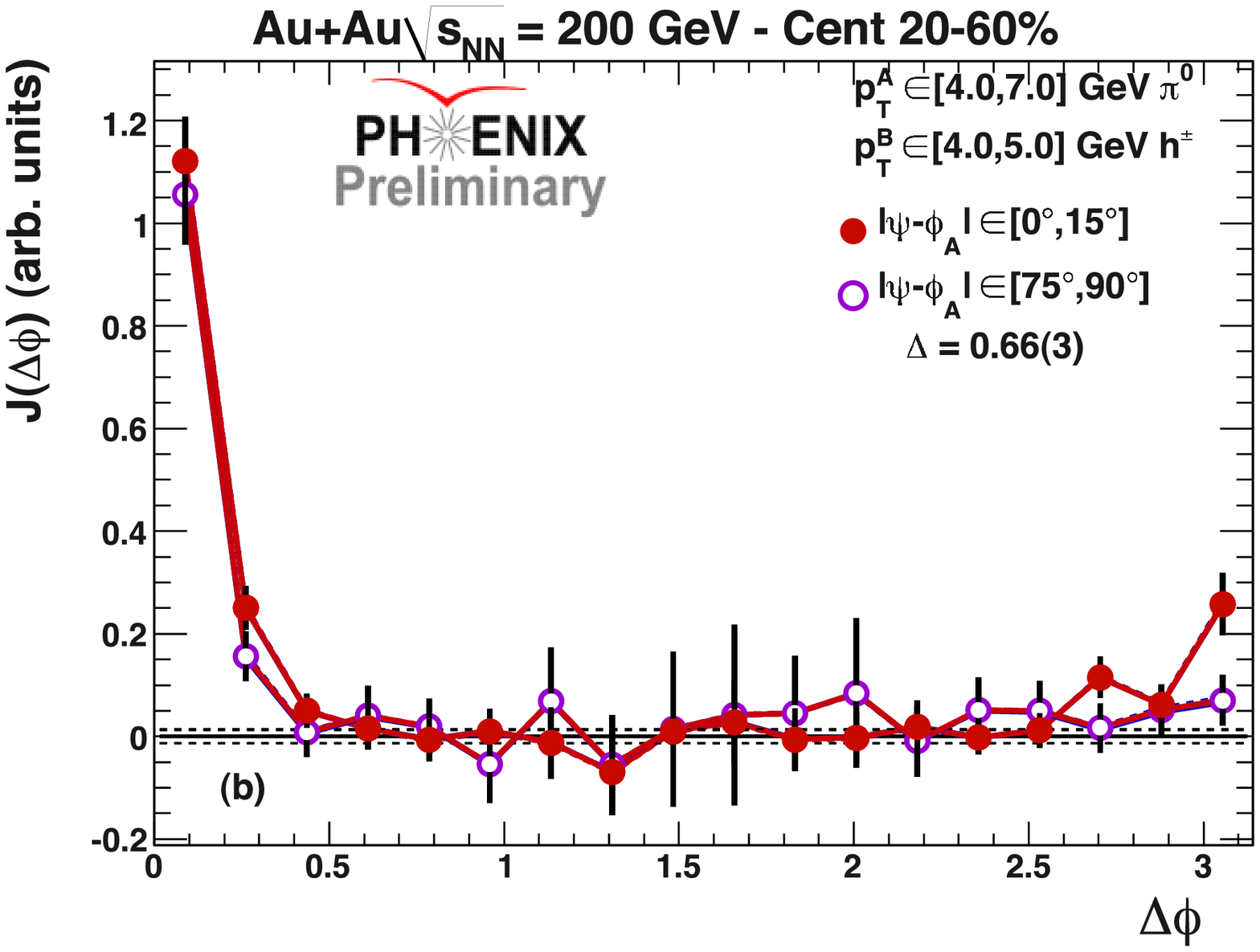}
  \includegraphics[width=1.0\linewidth]{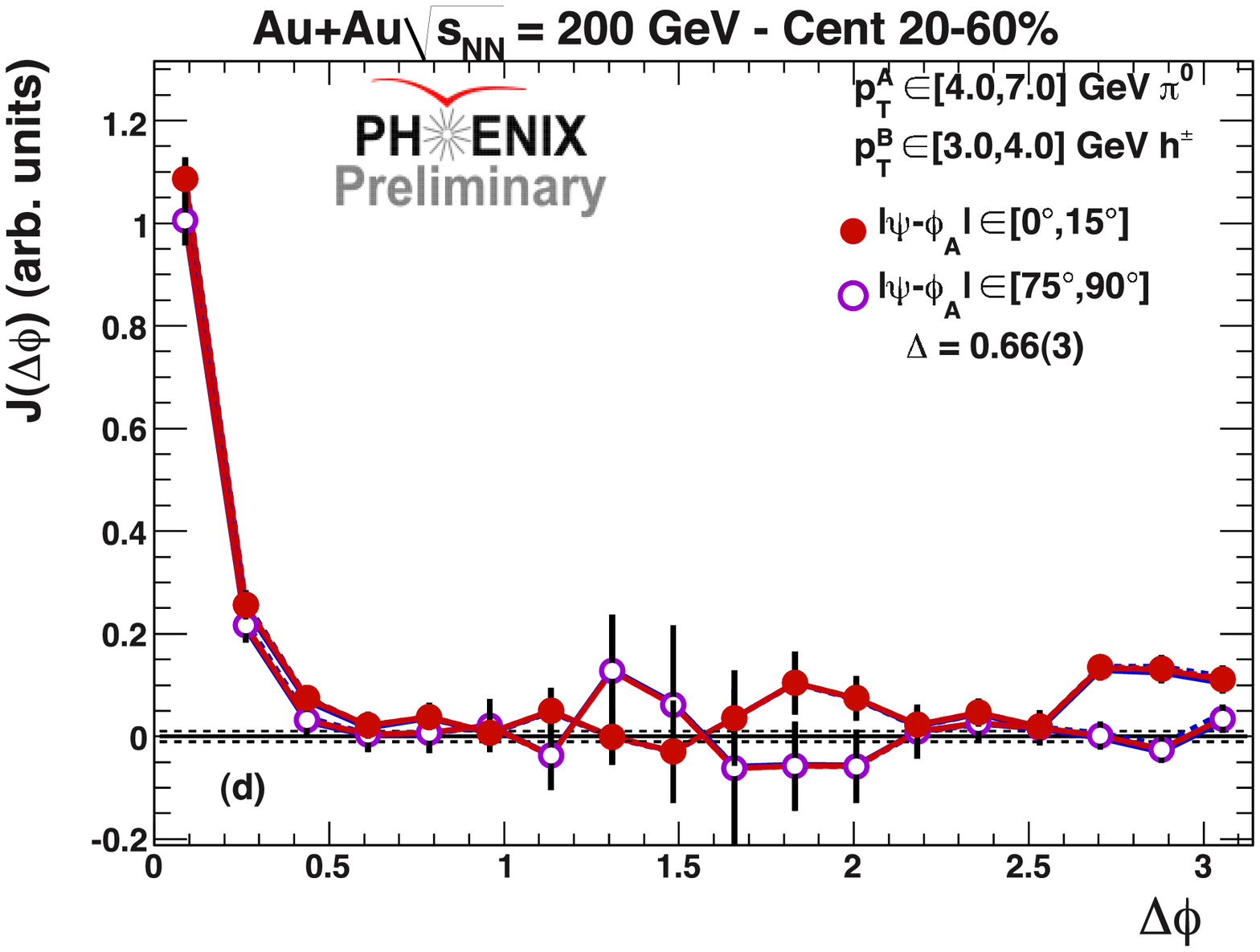}
\end{minipage}
\caption[]{
(Color online) Jet contributions in azimuthal angle are overlain 
for the most in-plane (solid symbols) and most out-of-plane (open 
symbols) selections.  Partner momenta, 4-5 GeV/c and 3-4 GeV/c, 
are shown in (a) \& (b) and (c) \& (d), respectively. Collision 
centralities 0-20\% and 20-60\% are shown in (a) \& (c) and (b) 
\& (d), respectively.
}
\label{dphi}
\end{figure}

Measured pairs are assumed to correlate trivially (two particles 
within the same event see the same reaction plane) or correlate 
via the same hard-scattering process. This two-source model 
assumption is expressed as \cite{ppg032}:
\begin{equation}
 C\left(\Delta\phi\right) = J\left(\Delta\phi)\right)+b_{0}\left(
1+\frac{\beta}{\alpha}\cos\left(2\Delta\phi\right)
+\frac{\gamma}{\alpha}\cos\left(4\Delta\phi\right)\right)
\label{eq1}
\end{equation}
 where the modulation variables from~\cite{flowsub} are:
\begin{eqnarray}
\alpha &=& 1 + 2 v^{A}_{2}cos\left(2\phi_{s}\right)
\frac{sin\left(2c\right)}{2c}\Delta 
+ 2 v^{A}_{2}cos\left(4\phi_{s}\right)\Delta_{4}
\nonumber \\
\beta &=& 2 v^{A}_{2} v^{B}_{2} + 2 v^{B}_{2} 
\left(1 + v^{A}_{4}\right)cos\left(2\phi_{s}\right)
\frac{sin\left(2c\right)}{2c}\Delta 
\nonumber \\
&+& 2 v^{A}_{2} v^{B}_{2} cos\left(4\phi_{s}\right)
\frac{sin\left(4c\right)}{4c}\Delta_4 + 2 v^{B}_{2} v^{A}_{4} 
cos\left(6\phi_{s}\right)\frac{sin\left(6c\right)}{6c}\Delta_6
\nonumber \\
\gamma &=& 2 v^{A}_{4} v^{B}_{4} + 2 v^{B}_{4} \left(1 + v^{A}_{2}\right)
cos\left(4\phi_{s}\right)\frac{sin\left(4c\right)}{4c}\Delta_4
\nonumber \\
&+& 2 v^{A}_{2} v^{B}_{4} \left( cos\left(2\phi_{s}\right)
\frac{sin\left(2c\right)}{2c}\Delta + cos\left(6\phi_{s}\right)
\frac{sin\left(6c\right)}{6c}\Delta_6 \right) 
\nonumber \\
&+& 2 v^{A}_{4} v^{B}_{4} cos\left(8\phi_{s}\right)
\frac{sin\left(8c\right)}{8c}\Delta_8
\label{eq2}.
\end{eqnarray}
These account for the reaction plane trigger binning and resolution
effects on the underlying event shape. Taking $c=\pi/2$, $\phi_{s}=\pi/4$,
and truncating higher-order terms, this form reduces to the 
$v^A_{2}v^B_{2}$ modulation used in previous reaction plane summed 
results such as those in \cite{ppg083}. Given sufficient detector 
resolution and narrowness of binning, the elliptic term to 
the underlying event modulation will flip sign between in-plane 
and out-of-plane bins as shown in Fig \ref{cfs}.
Examples of the resulting subtracted jet distributions are shown 
in Fig~\ref{dphi}. 

Elliptic ($v_{2}$) and hexadecapole ($v_{4}$) terms are measured 
for both trigger and partner particles via Fourier decomposition 
of the singles distributions with respect to the reaction plane. 
The systematic uncertainties on the underlying event modulation 
are of two kinds: those that correlate with $\phi_{s}$ and those 
that anti-correlate with $\phi_{s}$. Uncertainties of both kinds 
are propagated and found to be negligible for these transverse 
momenta.

The underlying event normalization ($b_{0}$) has been determined 
via the Zero Yield at Minimum (ZYAM) assumption. In the procedure, 
the correlations are fit to a functional form containing a near-side 
Gaussian, an away-side Gaussian, and a flow term as described
above. The well-separated near-side and away-side jet production 
gives a broad region over which the underlying event contribution 
dominates. Normalization uncertainty from the ZYAM procedure is
estimated via a Monte Carlo using the measured statistical precision 
as input. The extracted uncertainty is small, but fully correlated 
along $\Delta\phi$. These values are depicted as dashed bands 
about 0 on the subtracted jet functions shown in Fig~\ref{dphi}.

Near- and away-side PTYs are found by integrating across $\Delta\phi$ 
windows that approximate $2\sigma$ jet-widths as measured from the 
reaction plane summed distribution.  The integration windows are 
$\Delta\phi~\epsilon~[7/9\pi,\pi]$ and $\Delta\phi~\epsilon~[15/18\pi,\pi]$ 
for 3-4 GeV/c and 4-5 GeV/c partners, respectively. This choice 
is made to reduce the effect of decreased precision near $\Delta\phi
= 90^{\circ}$ and the influence of the $\Delta\phi$-correlated ZYAM 
uncertainty. These windows introduce little bias as only small 
amounts of jet yield remain outside the selections.

\begin{figure}[tb]
\begin{minipage}{0.5\textwidth}
  \includegraphics[width=1.0\linewidth]{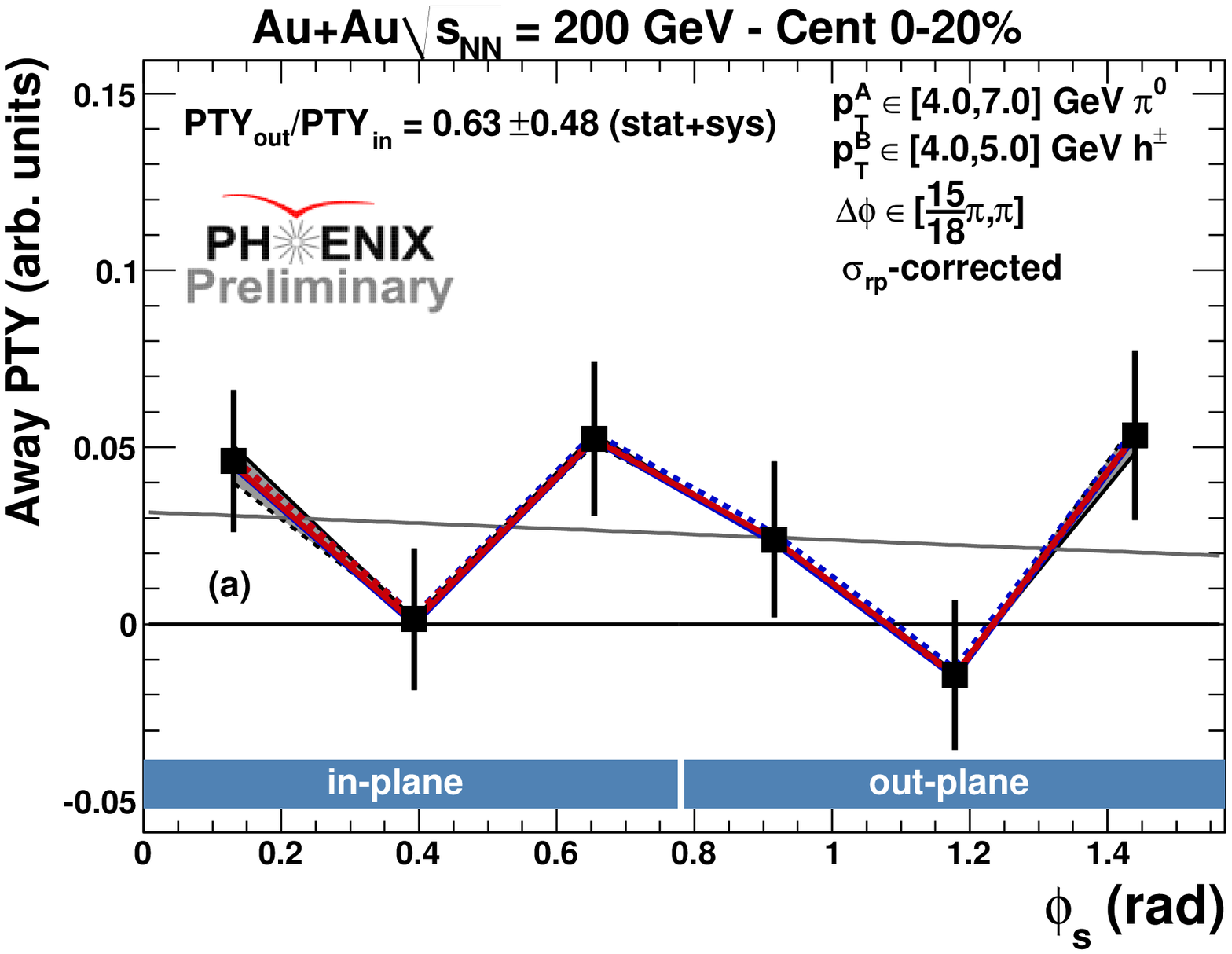}
  \includegraphics[width=1.0\linewidth]{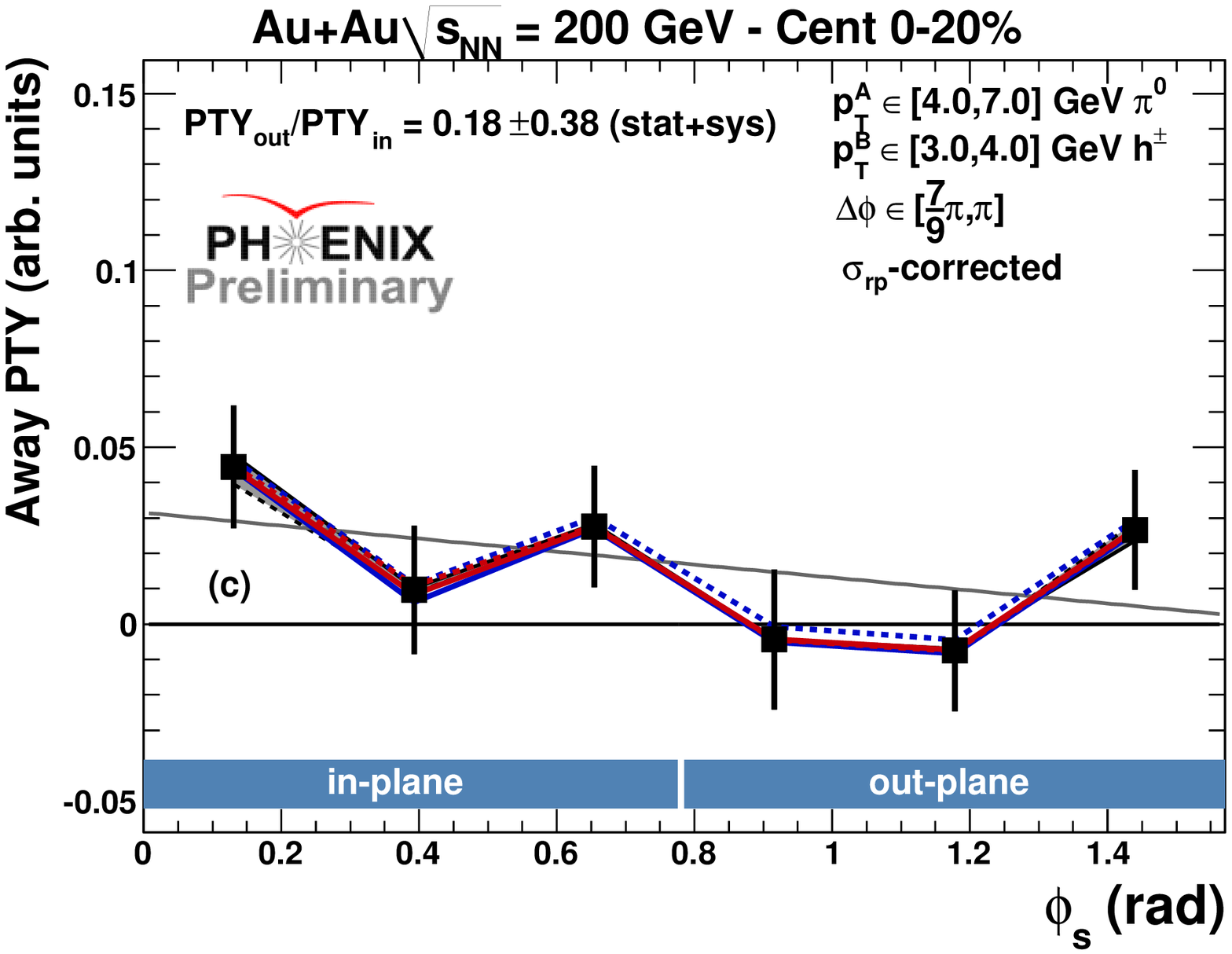}
\end{minipage}
\begin{minipage}{0.5\textwidth}
  \includegraphics[width=1.0\linewidth]{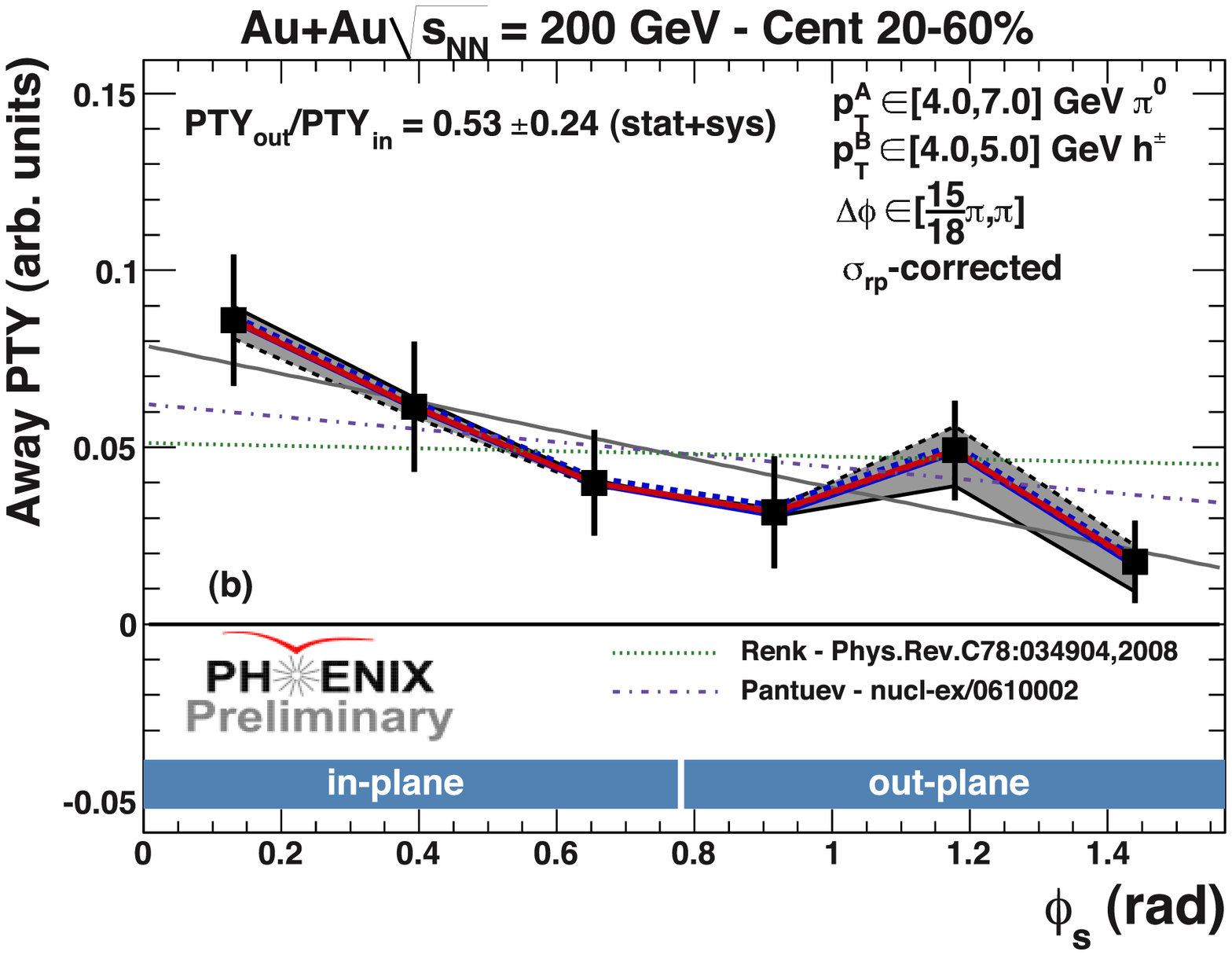}
  \includegraphics[width=1.0\linewidth]{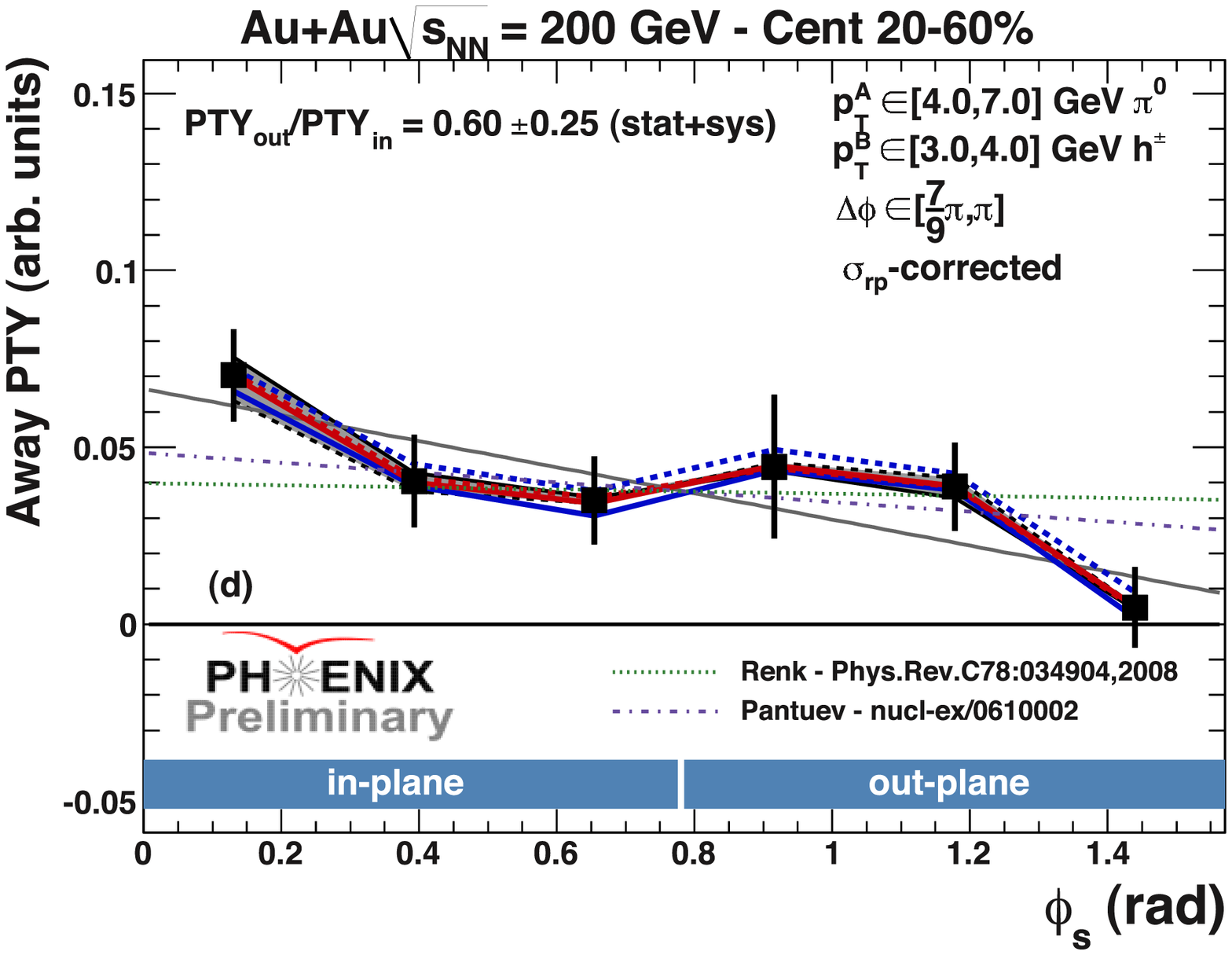}
\end{minipage}
\caption[]{
(Color online) Integrated away-side jet contributions along angle 
with respect to the reaction plane.  Partner momenta, 4-5 GeV/c
and 3-4 GeV/c, are shown in (a) \& (b) and (c) \& (d), respectively.
Collision centralities 0-20\% and 20-60\% are shown in (a) \& (c) 
and (b) \& (d), respectively.
}
\label{phis}
\end{figure}

Near- and away-side integrated PTY $\phi_{s}$ are shown in Fig~\ref{phis}. 
These distributions have been corrected for the smearing caused 
by the reaction plane resolution as was done in \cite{rpunsmear}.
In general the corrections are small and trends appear in the 
raw data as can be seen already in Fig \ref{dphi}. The uncertainty 
from the resolution correction is shown as an additional systematic 
band and is fully anti-correlated along $\phi_{s}$.
 
\section{Results}\label{results}

Subtracted jet functions for the most in-plane and the most 
out-of-plane $15^{\circ}~\phi_{s}$ bins are shown in Fig~\ref{dphi}. 
The jet functions for each partner momenta share a common arbitrary 
vertical scale. The near-side PTYs for both partner momenta in 
the most central and mid-central collisions are consistent within 
$\pm1\sigma$ to a flat $\phi_{s}$ dependence. The mid-central 
away-side PTYs, shown in Fig~\ref{phis}, exhibit a steeply falling 
trend with reaction plane angle.  More central away-side PTYs 
lack statistical sensitivity to differentiate between rising and 
falling trends.

\begin{figure}[tb]
  \centering
  \includegraphics[width=0.667\textwidth]{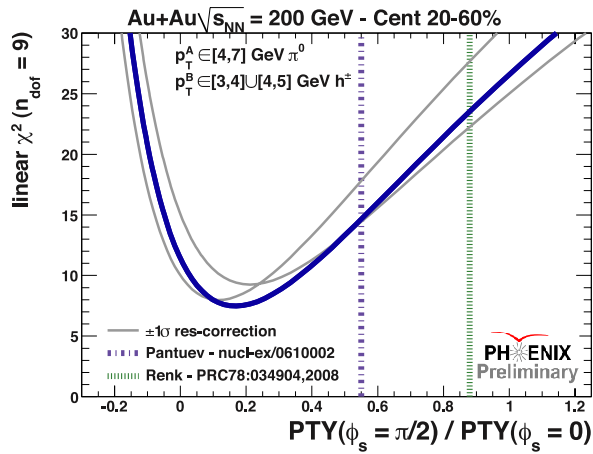}
\caption[]{
(Color online) $\chi^{2}$ distribution of away-side suppression 
with respect to the reaction plane reported as the ratio of the 
PTYs at the extrema of most in-plane and most out-of-plane. Fits 
to $\pm1\sigma$ resolution correction uncertainty are also reported.
Increasing suppression trends with angle relative to the reaction 
plane lie between 0-1. Falling suppression trends with respect 
to the reaction plane lie above 1.
}
\label{chi2}
\end{figure}

Composite $\chi^{2}$ distribution for linear fits to both partner 
momenta under the assumption of identical percentage variation 
in the away-side PTYs along $\phi_{s}$ for both partner momenta is 
shown in Fig~\ref{chi2}.
Percentage variations are reported as the ratio of PTY at the 
extrema of most in-plane and most out-of-plane. A $\chi^{2}/dof$ 
of 0.8 for 9 degrees of freedom is found for the best fit to the 
data. The gray lines show the agreement of fits made against the 
$\pm1\sigma$ reaction plane resolution corrections. These uncertainties 
do not correlate with momentum selection. The data rule out rising 
variations (values above 1) to more than 4$\sigma$. The data also 
prefer more strongly falling variations and do not rule out complete 
suppression for $\phi_{s} = 90^{\circ}$. Renk's prediction, made 
for higher $p_{T}$ than these measurements, is incompatible with 
these data at lower momenta~\cite{renk}. The prediction from Pantuev 
is a closer match but still shows a weaker dependence than these 
data~\cite{vlad}.

\section{Conclusions}\label{concl}

We have shown that away-side suppression increases with increasing 
angle with respect to the reaction plane in mid-central Au+Au 200 
GeV collisions. This dependence shows that the surviving back-to-back 
pairs in mid-central collisions originate predominantly from partons 
that have crossed some portion of the nuclear overlap and not from 
unmodified partons emitted tangentially to the overlap surface. The 
current central measurements do not constrain which category of 
production dominants in central events.  As spectral slopes of 
back-to-back high $p_{T}$ PTYs are known not to change in mid-central 
collisions from p+p values~\cite{ppg083}, the data increasingly 
support a picture where a fraction of partons in mid-central 
collisions are crossing the nuclear overlap with very little 
energy loss. Lumpiness in the event-to-event sampling of the initial
state distribution of soft production may play a role in back-to-back
jet survival through the nuclear overlap.

The steep dependence measured presents a challenge for theories 
that predict only a weak dependence of away-side suppression with 
respect to the reaction plane. Many theories that otherwise describe 
the overall level of nuclear suppression of both singles and pairs 
in heavy ion collisions may be able to describe this data by 
implementing initial state distributions for soft production that 
contain additional anisotropy. Thus these data may play an important 
roll in further constraining the geometry of the initial state 
in heavy ion collisions. The geometric distribution of soft production 
is a topic of considerable interest as it is an important input 
for the determination of medium viscosity from flow measurement 
descriptions.

\bibliography{bigsky2009-template}
\bibliographystyle{bigsky2009}
 
\vfill\eject
\end{document}